\documentclass[amsmath,amssymb,onecolumn, a4paper,14pt]{revtex4}
\usepackage{amsmath}
\usepackage{amssymb}
\usepackage{amsmath,amsthm,amssymb,amscd}
\usepackage{latexsym}
\usepackage{indentfirst}
\usepackage{subfigure}
\usepackage{graphicx}
\usepackage{extarrows}
\usepackage{bm}
\usepackage{multirow}
\usepackage[colorlinks,linkcolor=blue,hyperindex,dvipdfm]{hyperref}

\date{\today}

\begin{document}

\title{
  \bf Improvement of risk estimate on wind turbine tower buckled by hurricane
}

\author{Jingwei Li, Yunxin Zhang} 
\affiliation{Laboratory of Mathematics for Nonlinear Science, School of Mathematical Sciences, Fudan University, Shanghai 200433, China. }

\begin{abstract}
Wind is one of the important reasonable resources. However, wind turbine towers are sure to be threatened by hurricanes. In this paper, method to estimate the number of wind turbine towers that would be buckled by hurricanes is discussed. Monte Carlo simulations show that our method is much better than the previous one. Since in our method, the probability density function of the buckling probability of a single turbine tower in a single hurricane is obtained accurately but not from one approximated expression. The result in this paper may be useful to the design and maintenance of wind farms.
\end{abstract}


\maketitle

\section{Introduction}

There are rich offshore wind resources all over the world. Wind is the one with the largest installed-capacity growth from 2007 to 2012 among all renewable resources in America. U.S. wind power capacity increased from 8.7 GW in 2005 to 39.1 GW 2010 \cite{DOE2010}. The National Renewable Energy Laboratory (NREL) estimates that offshore wind resources can be as high as four times the U.S. electricity generating capacity in 2010 \cite{Shwartz2010}.

Although the offshore wind resources are great, it is necessary to foresee the hurricane risks to offshore wind turbines. U.S. offshore resources are geographically distributed through the Atlantic, Pacific and Great Lake coasts. The most accessible shallow resources are located in the Atlantic and Gulf Coasts. Resources at depths shallower than 60 m in the Atlantic coast, from Georgia to Maine, are estimated to be 920 GW; the estimate for these resources in the Gulf coast is 460 GW \cite{Shwartz2010}. Offshore wind turbines in these areas will be at risk from Atlantic hurricanes. Between 1949 and 2006, 93 hurricanes struck the U.S. mainland according to the HURDAT (Hurricane Database) database of the National Hurricane Center \cite{Blake2007}. Hurricane risks are quite variable, both geographically and temporally. Pielke {\it et al.} note pronounced differences in the total hurricane damages (normalized to 2005) occurring each decade \cite{Pielke2008}. Hurricanes and other hazards can cause widespread electric power outages, which, in turn, can affect business
operations, heating, financial transactions, security systems,
water distribution, traffic signaling, and countless other aspects
of daily life. And hurricane hazards cause the most damage to
power systems in the Eastern U.S \cite{Liu2007}. Interdecadal major hurricane fluctuations occur in both landfall locations and overall activity \cite{Herbert1996,Landsea1992,Gray1997,Landsea1999}. Most of the deadliest and costliest Atlantic tropical cyclones are major hurricanes \cite{Herbert1996}. Major hurricanes account for just over 20\% of the tropical storms and hurricanes that strike the United States but cause more than 80 percent of the damage \cite{Landsea1998}.

Hurricanes Katrina and Rita hit the center of the American petrochemical industry, shutting down eight refineries, hundreds
of oil-drilling and production platforms, and numerous other industrial facilities. Furthermore, they triggered numerous hazardous-materials (hazmat) releases from industrial facilities and storage terminals onshore, as well as from oil and gas production
facilities offshore in the Gulf of Mexico (GoM) \cite{Cruz2009}. Hurricane Ivan had caused much
concern among industrialists, operators, and government officials
on the performance of the offshore oil and gas drilling and
production activities and infrastructure in the GoM during major
hurricanes \cite{Ward2005}. Hurricanes in the GoM, which can reach wind speeds of
240 km/h accompanied by waves of over 25 m, pose serious
challenges for the design and operation of offshore facilities in
these harsh environments \cite{Leffler2003}.

The 2005 Atlantic hurricane season was the most active hurricane season on record with 28 named storms (previous record was 21 named storms set in 1933), 15 of which reached
hurricane status (previous record was 12 set in 1969) \cite{NHC2006}. The IPCC suggests in 1990
that: there are some evidences from model simulations and empirical considerations that the frequency per year, intensity and area of occurrence of tropical disturbances may increase in a doubled carbon dioxide world, though it is not compelling \cite{Houghton1990}. Recent sizable hurricane losses have raised questions
about the causes. Some have claimed that storm frequencies and/or intensities have
increased \cite{LandseaC1999,Kunkel2008}, but other studies indicate no long-term trend
in hurricane activity during the 20th century \cite{LandseaC2005}. Others see the increases as a
result of climate change resulting from global warming \cite{Emanuel2005,Emanuel2008}. Others consider the higher hurricane losses a result of societal changes leading to
greater vulnerability in hurricane prone areas \cite{Pielke2005}.

The average annual insured losses from hurricanes are 2.6 billion dollars for 1949-2006, the highest storm-related loss in the nation. The ranks of the average annual
insured property losses of all the extreme weather conditions in the U.S. (2006 dollars in
billions) are: (1) hurricanes (2.6 billion dollars), (2) floods (2.2 billion dollars), (3) thunderstorms (1.6 billion dollars), (4) tornadoes
(1.0 billion dollars), (5) hail (0.9 billion dollars), (6) snowstorms (0.5 billion dollars), (7) freezing rainstorms (0.2 billion dollars), and (8)
wind storms (0.2 billion dollars) \cite{Changnon2001}.  On the other hand, world energy problems are becoming more and more serious. It is imminent to make use of rich wind resources. While the development of onshore wind turbines has gone on wheels, there are 20 offshore wind projects in the planning process \cite{Shwartz2010}.

So it is important to analyze the hurricane risks to offshore wind turbines. The design, maintenance and assessment of offshore structures
must meet the requirements laid down in the Code of Federal
Regulations, Title 30, Part 250 \cite{30CFR2502006}. And the design requirements
for offshore facilities follow the recommendations of the
American Petroleum Institute which calls for platforms and
floating permanent systems to be designed to withstand a full-population
hurricane \cite{Wisch2004} with a return period of
100 years \cite{API1997,API2000}. These 100-year criteria correspond to
a wind speed of about 150 km/h in 1-h average winds or about
180 km/h in sustained 1-min winds and a maximum wave height
of 22m \cite{Ghonheim2005,Kramek2006,Ward2005}.

In \cite{Stephen2012}, a series of models used to describe the characters of hurricanes and turbine towers are established, based on which the risks on offshore wind turbines from hurricanes are discussed, and a model used to estimate them in four representative locations (Galveston County, TX; Dare County, NC; Atlantic County, NJ; and Dukes County, MA) in the Atlantic and Gulf Coastal waters of the United States has been built. 
Although the models in \cite{Stephen2012} can already give a good and reasonable risk estimate of hurricanes, one approximated expression used to calculate the probability density function of turbine buckling probability is able to be improved. In this study, we will give out an accurate expression of it. Some discussions and results given in \cite{Stephen2012} are based on Monte Carlo simulations. One of them has its mathematical expression which will be discussed later. Thanks to this work, Monte Carlo simulations can be avoided. Perhaps it will be helpful for further theoretical analyses. At last, we will put forward an idea of estimating hurricane risks which depends on the expected survive time (EST).

The organization of this paper is as follows. The basic models and the improved methods will be briefly described in the next section, and then in Sec. {\bf III} the analyses on Monte Carlo simulations will be presented. In sec. {\bf IV}, a short discussion about expected survive time (EST) will be given. The results will be given in Sec. {\bf V}, and then concluding remarks in the final section.


\section{Models of hurricanes and turbines}
\subsection{The models established in \cite{Stephen2012}}

The previous model presented by Stephen {\it et al.} in \cite{Stephen2012} can be summarized as follows.

\noindent{\bf(1) Hurricane occurrence.}

Hurricane occurrence is modeled as a Poisson process with rate $\lambda$ obtained by fitting to historical hurricane data. The probability that $H$, the number of hurricanes that occur in $T$ years, equals a particular value $h$ is given by:
\begin{equation}
Pr(H=h)=\frac{(\lambda T)^h}{h!}e^{-\lambda T}.\label{poisson}
\end{equation}

\noindent{\bf(2) The maximum 10-min sustained wind speed of each hurricane at 10-m height.}

 It is supposed that there is a maximum 10-min sustained wind speed during each hurricane. The wind speed at 90-m height (hub-height of turbine towers) decides the probability of a single wind turbine tower buckling \cite{IEC}. The maximum 10-min sustained wind speed of each hurricane at 10-m height is modeled as Generalized Extreme Value (GEV) distribution with a location parameter $\mu$, a scale parameter $\sigma$, and a shape parameter $\xi$ fitted to historical hurricane data. The probability density function for $W$, the maximum sustained wind speed, evaluated at particular value $w$ is given by:
\begin{equation}
f_W(w)=\frac {1}{\sigma} \exp\left(-\left(1+\xi \frac{w-\mu}{\sigma}\right)^{- \frac {1}{\xi}}\right) \left(1+\xi \frac {w-\mu}{\sigma}\right)^{-1- \frac{1}{\xi}}.\label{GEV}
\end{equation}

\noindent{\bf(3) The buckling probability of a single wind turbine tower.}

The probability that a single wind turbine tower is buckled by a maximum 10-min sustained hub-height wind speed $u$ is modeled using a log-logistic function with a scale parameter $\alpha$ and a shape parameter $\beta$. These parameters are fitted to probabilities of turbine tower buckling calculated by comparing the results of simulations of the 5-MW offshore wind turbine designed by the NREL to the stochastic resistance to buckling proposed by S{\o}rensen, et al. \cite{Jonkman2009,Sorensen2005}. The buckling probability of a single wind turbine $B$ can be given by the following log-logistic function:
\begin{equation}
B(u)=\frac{{\left(\frac {u}{\alpha}\right)}^{\beta}}{1+{\left(\frac {u}{\alpha}\right)}^{\beta} }.\label{loglogistic}
\end{equation}
In order to distinguish the sign of buckling probability from the differential sign, we use $B$ to express the random variable and $b$ to express its exact value instead of $D$ and $d$ used by Stephen {\it et al.} originally.

\noindent{\bf(4) Fitting to Beta distribution.}

The exact hub-height (90-m) wind speed $u$ is related to the exact value $w$ of 10-m height random wind speed $W$. This wind speed is scaled from 90-m height to 10-m height assuming
power-law wind shear with an exponent of 0.077 \cite{Franklin2003}. It can be expressed as:
\begin{equation}
0.077=\frac{\ln(u/w)}{\ln(90/10)}.\label{windshear}
\end{equation}

A Beta distribution is fitted to Monte Carlo simulations of the convolution of $B$ and $W$. The procedure used in \cite{Stephen2012} is as follows:\\
{\bf 1.} Simulate a large number of wind speeds $W$ with a Generalized Extreme Value (GEV) distribution in Eq. (\ref{GEV}).\\
{\bf 2.} Calculate the probability of turbine tower buckling $B$ for each wind speed from  {\bf step. 1} using the log-logistic damage function in Eq. (\ref{loglogistic}) and the wind-shear function in Eq. (\ref{windshear}).\\
{\bf 3.} Calculate the empirical cumulative distribution function (CDF) for the buckling probabilities in {\bf step. 2}. For each probability of buckling, calculate the probability that value occurs. The result is an $x$-$y$ graph with {\bf Probability of turbine tower buckling} on the $x$-axis and {\bf Probability of occurrence} on the $y$-axis.\\
{\bf 4.} Use nonlinear curve fitting to fit the CDF of a Beta distribution to the empirical CDF in {\bf step. 3}. Use starting values of $\alpha_B = 0.02$ and $\beta_B = 0.2$.

The probability density function (PDF) of beta distribution with parameters $\alpha_B$ and $\beta_B$ is given by:
\begin{equation}
f(x;\alpha_B,\beta_B )=\frac {x^{\alpha_B-1} (1-x)^{\beta_B-1}}{\int_0^1 u^{\alpha_B-1} (1-u)^{\beta_B-1} du}.\label{betadistribution}
\end{equation}

\noindent{\bf(5) The probability of buckling turbine number.}

Through Eq. (\ref{betadistribution}), it is easy to show that the number of turbines buckled by a single hurricane in a wind farm with $n$ turbines can be modeled by a beta-binomial distribution with parameters $\alpha_B$ and $\beta_B$. The probability that $X$, the number of turbine towers that buckle out of $n$ total, equals a particular value $x$ is given by:
\begin{equation}
Pr(X=x)={n \choose x}  \frac{B(x+\alpha_B,n-x+\beta_B )}{B(\alpha_B,\beta_B )}.\label{betabinomialdistribution}
\end{equation}

\noindent{\bf(6) The number of buckling turbines in time period T.}

The cumulative distribution of the buckling turbine number in $T$ years without replacement, $Y_{\rm{no\ rep}}$, is modeled as a modified phase-type distribution \cite{Neuts1995,BladtM2005}:
\begin{equation}
Pr(Y_{\rm{no\ rep}}\le y | \tau \le t)={\bf g} \exp(T{\bf T}(y,n)) {\bf e}.\label{phasetype}
\end{equation}
where ${\bf g}$, ${\bf e}$ are vectors, and ${\bf T}$ is one matrix (see \cite{Stephen2012} for detailed explanation for them).

\subsection{Our improved method}

{\bf Step. (1)-(3)} are the methods to establish basic mathematical models of hurricanes and turbine towers. {\bf Step. (4)-(6)} are the methods to solve the models.

Monte Carlo simulations are used to estimate the CDF of the buckling probability of a single turbine tower in a single hurricane through the procedure presented in {\bf step. (4)}. In fact, the buckling probability for a single turbine in a hurricane $B$ is a function of maximum 10-min sustained wind speed $u$ at hub-height, while $u$ and $w$ fit Eq. (\ref{windshear}), so its PDF can be calculated from the PDF of $W$ which is given by Eq. (\ref{GEV}). Here we need a theorem:

\noindent{\bf Theorem}
\label{ProOfFunc}
if $Y=f(X)$ is a monotonic function, $X=f^{-1}(Y)\in C^1(R)$, and the PDF of $X$ is $p(x)$, then the PDF of $Y$, $q(y)$ is given by:
\begin{equation}
q(y)=p(f^{-1}(y))|\frac{df^{-1}(y)}{dy}|.\label{qy}
\end{equation}
$\blacksquare$

By taking use of the above theorem, we can get the PDF of $B$ noted as $p(b)$:
\begin{equation}
p(b)=\frac{1}{\sigma}\exp(-(1+\xi\frac{\frac{\alpha}{s}(\frac{b}{1-b})^{\frac{1}{\beta}}-\mu}{\sigma})^{-\frac{1}{\xi}})
(1+\xi\frac{\frac{\alpha}{s}(\frac{b}{1-b})^{\frac{1}{\beta}}-\mu}{\sigma})^{-1-\frac{1}{\xi}}
\frac{1}{s}\frac{\alpha}{\beta}\frac{b^{\frac{1}{\beta}-1}}{(1-b)^{\frac{1}{\beta}+1}},\label{pb}
\end{equation}
where it is supposed that $u=sw$. $s=\exp(0.077\log(9))$ can be got from Eq. (\ref{windshear}).

In this paper, Eq. (\ref{pb}) will be used to estimate the number of buckling turbines in time period $T$ directly. Let $A(b)$ be one $(n+1)\times (n+1)$ matrix with its elements given by:
\begin{equation}\label{Ad}
A(b)(i,j)=\left\{\begin{array}{cc}
{n-i+1 \choose j-i}b^{j-i}(1-b)^{n-j+1}&i\le j\\
0&i>j
\end{array}\right.,
\end{equation}
where $b$ is the buckling probability of a single wind turbine after a single hurricane and $n$ is the number of turbines in one farm. We also define one state vector ${\bf f}=(f_1,f_2,f_3,\cdots,f_{n+1})$ with $f_i$ being the probability that $i-1$ turbines are buckled out of $n$ total after a single hurricane. ${\bf f}={\bf g}=(1,0,0,\cdots,0)$ at first, since none turbine is buckled. It is obvious that the state vector ${\bf f}={\bf g}\prod_{j=1}^kA(b_j)$ after $k$ hurricanes, where $b_j$ is a sample of $b$ in the $j$th hurricane. According to Eq. (\ref{poisson}), the probability that $k$ hurricanes happen in $T$ years is $\frac{(\lambda T)^k}{h!}e^{-\lambda T}$. The expected value $E({\bf f})$ of ${\bf f}$ can be obtained as follows:
\begin{equation}
E({\bf f})=E\left(e^{-\lambda T}+\sum_{k=1}^{\infty}{\bf g}\prod_{j=1}^kA(b_j)\frac{(\lambda T)^k}{h!}e^{-\lambda T}\right),\label{Ef1}
\end{equation}
where $b_j$ ($j=1,2,\cdots$) are independent of each other according to the assumption in \cite{Stephen2012} that the maximum 10-min sustained wind speed of each hurricane is independent of each other. So $E(\prod_{j=1}^kA(b_j))=\prod_{j=1}^kE(A(b_j))$. Notice that $E(A(b_i))=E(A(b_j))=E(A(b))$, since $b_j$ $(j=1,2,\cdots)$ are samples of $b$. Eq. (\ref{Ef1}) is equivalent to:
\begin{equation}
E({\bf f})={\bf g}\exp(\lambda TE(A(b)))e^{-\lambda T}={\bf g}\exp(\lambda T(E(A(b))-I)).\label{Ef2}
\end{equation}
Compare it with the expression ${\bf g} \exp(T{\bf T}(y,n))$ of ${\bf f}$ in Eq. (\ref{phasetype}), they are same except replacing the matrix ${\bf T}$ in Eq. (\ref{phasetype}) by $\lambda(E(A(b))-I)$. It can be expressed as follows:
\begin{equation}
E\left({n-i+1 \choose j-i}b^{j-i}(1-b)^{n-j+1}\right)\to {n-i+1 \choose j-i}\frac{B(j-i+\alpha_B,n-i+1+\beta_B )}{B(\alpha_B,\beta_B)},\label{ess}
\end{equation}
where $B()$ is the beta function in Eq. {\ref{betabinomialdistribution}}. And we have:
\begin{equation}
E\left({n-i+1 \choose j-i}b^{j-i}(1-b)^{n-j+1}\right)=\int_{a_1}^{a_2}{n-i+1 \choose j-i}b^{j-i}(1-b)^{n-j+1}p(b)db.\label{Aconp}
\end{equation}
$a_1$ and $a_2$ are the bounds of $b$ which are involved by the GEV distribution of $W$. As mentioned before, $p(b)$ is directly used in this paper. Since the essence of this method is given out by Eq. (\ref{ess}), we can use it to calculate $Y_{\rm{rep}}$ which is a estimate of the buckling turbine number in $T$ years with replacement after each hurricane (see \cite{Stephen2012} for detailed explanation for it).

It will be seen in Sec. {\bf V} that this method gives a closer estimate to the Monte Carlo simulations in both cases with or without replacement.

\section{The analysis on Monte Carlo simulations}

In \cite{Stephen2012}, the loss caused by the Category (a classification standard of hurricanes) 1 to 3 hurricanes is estimated by doing Monte Carlo simulations and excluding the simulations with Category 4 to 5 hurricanes happening manually, which is equivalent to replace the PDF of $B$ by the condition PDF of $B$ for $b\le b_{\rm{3max}}$. $b_{\rm{3max}}$ is the maximum buckling probability caused by a Category 3 hurricane of a single turbine tower. By Eq. (\ref{loglogistic}) and  Eq. (\ref{windshear}) one can get:
\begin{equation}
b_{\rm{3max}}=\frac{{\left(\frac {sw_{\rm{3max}}}{\alpha}\right)}^{\beta}}{1+{\left(\frac {sw_{\rm{3max}}}{\alpha}\right)}^{\beta} },\label{b3max}
\end{equation}
where $w_{\rm{3max}}$ is the largest value of maximum 10-min sustained wind speed at 10-m height of Category 3 hurricanes. $u=sw$ and $s=\exp(0.077\log(9))$ give the meaning of $s$. With the definition of $b_{\rm{3max}}$, we have the condition PDF of $B$ for $b\le b_{\rm{3max}}$ as:
\begin{equation}
p(b)|_{b\le b_{\rm{3max}}}=\frac{p(b)}{\int_{a_1}^{b_{\rm{3max}}}p(b)db}.\label{PdfBCond}
\end{equation}
It will be seen in Sec. {\bf V} that this method fits the Monte Carlo simulations very well.

\section{Expected survive time (EST)}

In this section, one new way will be given to analyze the risks of a wind farm suffered from hurricanes. It is supposed that the buckling of each turbine is independent of each other. If the character of an individual turbine buckling is confirmed, it can describe the character of the whole wind farm in some aspects. The probability $q$ of one turbine {\bf not} buckling in $t_0$ years is given by:
\begin{equation}
q=E(e^{-\lambda t_0}+\sum_{k=1}^{\infty}\prod_{j=1}^{k}(1-b_j)\frac{(\lambda t_0)^k}{h!}e^{-\lambda t_0}),\label{q1}
\end{equation}
which is equivalent to
\begin{equation}
q=e^{-\lambda t_0E(b)}.\label{q2}
\end{equation}
Here $b_j$ means the probability of this individual turbine buckling in $j$th hurricane. Naturally the probability of one turbine buckling in $t_0$ years, noted as $1-q$ is given by:
\begin{equation}
1-q=1-e^{-\lambda t_0E(b)}.\label{1q}
\end{equation}
Actually one turbine buckling in $t_0$ years means the exact value $t$ of its survive time $T$ is less than $t_0$. So $1-q$ is the CDF of $T$. The PDF of $T$, noted as $p(t)$ can be expressed as:
\begin{equation}
p(t)=\lambda E(b)e^{-\lambda tE(b)}.\label{pt}
\end{equation}
We have the expected survive time $E(T)$ as:
\begin{equation}
E(T)=\int_0^{\infty}tp(t)dt=\frac{1}{\lambda E(b)}.\label{ET}
\end{equation}
The expected survive time $E(T)$ means the mean value of survive time $T$. It is determined by $E(b)$ and $\lambda$ only. $E(b)$ and $\lambda$ are parameters which reflect the surrounding model of the wind farm. The worse is the surrounding, the less is $E(T)$. So $E(T)$ can be used to compare wind farms with each other or decide whether a project should be applied to a wind farm. We have $E(b)$ as:
\begin{equation}
E(b)=\int_{a_1}^{a_2}bp(b)db.\label{Eb}
\end{equation}
It is easy to show that:
\begin{equation}
1-q=1-e^{-\frac{T}{E(T)}}\label{1q2}
\end{equation}
by using Eq. (\ref{1q}) and Eq. (\ref{ET}). We can get the probability of an individual turbine buckling in $T$ years only with EST known. The expected survive number (ESN) $E(l)$ can be calculated by:
\begin{equation}
E(l)=n(1-q)=n(1-e^{-\frac{T}{E(T)}}),\label{El}
\end{equation}
where $n$ is the total turbine number. It will be seen in Sec. {\bf V} that ESN got by Eq. (\ref{El}) is close to the results of Monte Carlo simulations and the state vector calculated by the method introduced in Sec. {\bf II}.

\section{Results}

\begin{figure}
  \centering
  \includegraphics[width=12cm]{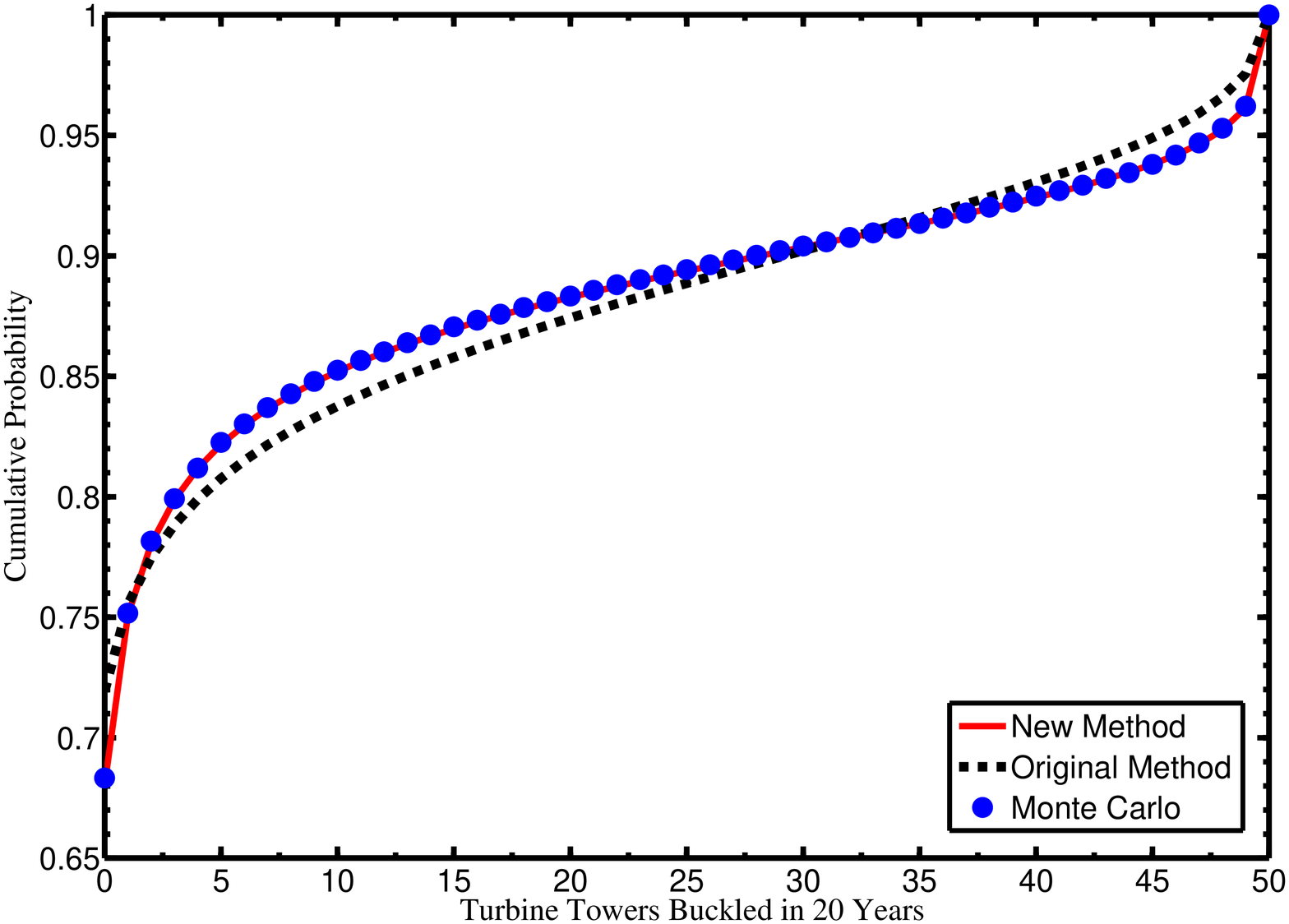}\\
  \caption{The cumulative distribution function (CDF) of buckling number of 50 turbines totally in 20 years. Buckling turbines aren't replaced. $x$-axis is the buckling number of wind turbine towers after 20 years. The value of $y$-axis at $x$ is the probability that the buckling number of wind turbine towers is less than $x$. Turbines are actively yawing in Galveston County, TX. The CDF is obtained by Eq. (\ref{pb}) with parameters $\alpha=174$, $\beta=19.3$, $\lambda=0.19$, $\mu=78.7$, $\sigma=12.1$, $\xi=0.251$.}\label{CompareOfYnorep}
\end{figure}

In our calculating, the same data and parameters in \cite{Stephen2012} are used (for details, see the captions of Figs. 1-3).
Firstly, our model is tested in calculating the CDF of buckling number without replacement in the wind farm of Galveston County, TX. Suppose that turbines are pointed into wind (Active Yawing). Test period $T$ and total turbine number $n$ are set to be 20 and 50 respectively. So are the following tests. We always use full line in red for the new method, dotted line in blue for Monte Carlo simulations and chain line in black for the original method in \cite{Stephen2012}. The result of new method is almost same to Monte Carlo simulations according to Fig. (\ref{CompareOfYnorep}), which shows the accuracy of the new method is better.

\begin{figure}
  \centering
  \includegraphics[width=12cm]{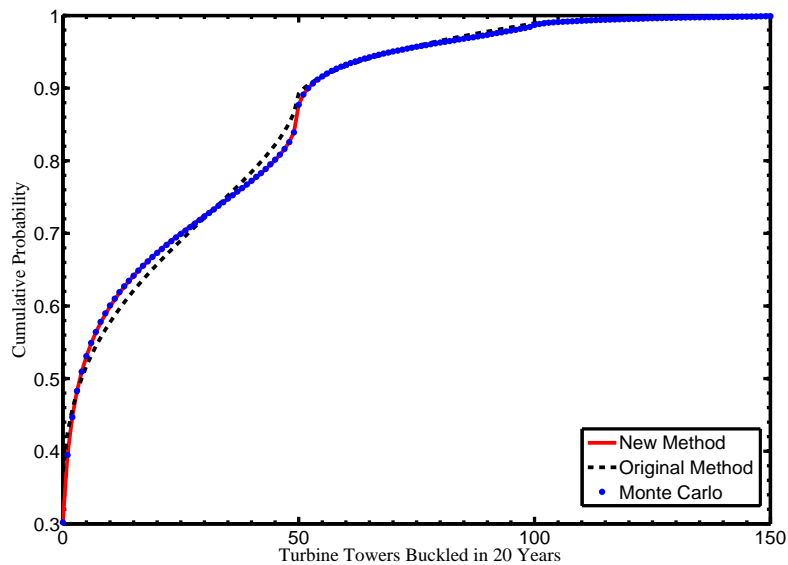}\\
  \caption{The cumulative distribution function (CDF) of buckling number of 50 turbines totally in 20 years. Buckling turbines are replaced after each hurricane. $x$-axis is the buckling number of wind turbine towers after 20 years. The value of $y$-axis at $x$ is the probability that the buckling number of wind turbine towers is less than $x$. Turbines are not yawing in Galveston County, TX. The CDF is obtained by Eq. (\ref{pb}) with parameters $\alpha=140$, $\beta=18.6$, $\lambda=0.19$, $\mu=78.7$, $\sigma=12.1$, $\xi=0.251$.}\label{CompareOfYrep}
\end{figure}

Secondly, the new model is tested when buckling turbines are replaced after each hurricane in the same wind farm (Galveston County, TX). But this time, turbines are pointed perpendicular to wind (Not Yawing). Again the new method is closer to Monte Carlo simulations according to Fig. (\ref{CompareOfYrep}).

In these two tests listed above, the new method gives more accurate results of the losses, which may be helpful for some further estimates and analyses on risks suffered from hurricanes.

\begin{figure}
  \includegraphics[width=12cm]{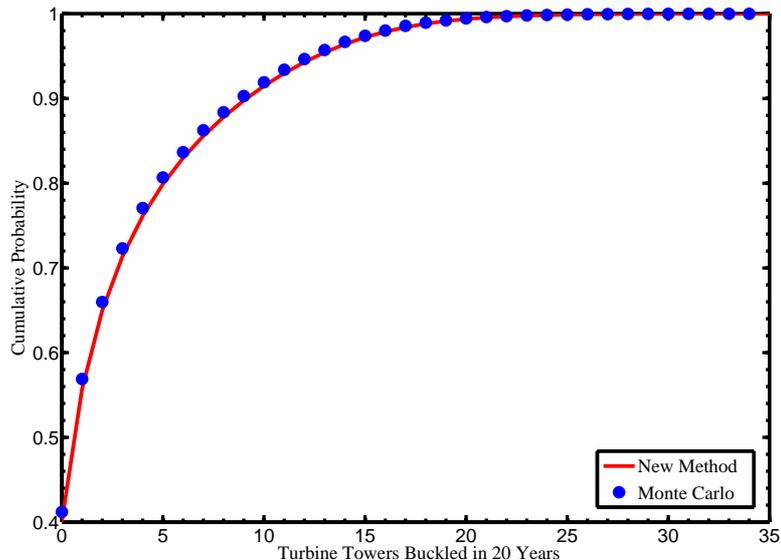}\\
  \caption{The cumulative distribution function (CDF) of buckling number of 50 turbines totally in 20 years without Category 4 to 5 hurricanes. Buckling turbines aren't replaced. $x$-axis is the buckling number of wind turbine towers after 20 years caused by Category 1 to 3 hurricanes. The value of $y$-axis at $x$ is the probability that the buckling number of wind turbine towers is less than $x$. Turbines are not yawing in Dare County, NC. The CDF is obtained by Eq. (\ref{PdfBCond}), Eq. (\ref{pb}) with parameters $\alpha=140$, $\beta=18.6$, $\lambda=0.21$, $\mu=77.6$, $\sigma=11.9$, $\xi=-0.0366$ and Eq. (\ref{b3max}) with parameters $w=113$.}\label{CompareOfYnorepCAT1to3}
\end{figure}

Then we test the condition PDF model mentioned in Sec. {\bf III} in the wind farm of Dare County, NC. Suppose that turbines are pointed perpendicular to wind (Not Yawing). The boundary value of 10-m wind speed between Category 3 and Category 4 hurricanes is not given out in \cite{Stephen2012}. So we choose it to be 113 knots. One can find in Fig. (\ref{CompareOfYnorepCAT1to3}) that the condition PDF of $B$ can give out a good estimate of the CDF of buckling number under a given condition, since its curve fits Monte Carlo simulations accurately.

Finally, the expected survive number (ESN) is calculated in three different ways. They are the state vector given by Eq. (\ref{Ef2}), Monte Carlo simulations based on Eq. (\ref{poisson}), Eq. (\ref{GEV}) and Eq. (\ref{loglogistic}), the expected buckling probability given by Eq. (\ref{1q}) respectively. The wind farm is chosen in TX and turbines are active yawing. These three results are extremely close to each other, which means that the buckling probability of a single turbine may reflect the risks of a wind farm suffered from hurricanes there reasonably.

As mentioned in Sec. {\bf III}, the mean of state vector in $T$ years without Category 4 to 5 hurricanes happening can be described easily by using a condition PDF of $B$. It is different from the situation that Category 4 to 5 hurricanes happen between Category 1 to 3 hurricanes while only the damage caused by Category 1 to 3 hurricanes is considered. It is easy to analyze it by Monte Carlo simulations. We have tried to give out the mathematical expression of the state vector caused by Category 1 to 3 hurricanes only when Category 1 to 3 hurricanes and Category 4 to 5 hurricanes happen alternately. However, it is difficult to handle this because of the correlations between hurricanes. This work should be meaningful since we can ensure the percentage of buckling turbines of each hurricane Category by the parameters of the surrounding in a wind farm [such as $\lambda$ in Eq. (\ref{poisson})] and the turbine [sucn as $\alpha$ and $\beta$ in Eq. (\ref{loglogistic})] directly but not by Monte Carlo simulations.

In Sec. {\bf IV}, the buckling probability of a single turbine in $T$ years is proposed. We notice that the buckling of each turbine in each hurricane is independent of each other. So we try to get the state vector of turbine buckling number after $T$ years by using Eq. (\ref{1q2}). However, the turbines in a same wind farm are in fact not independent, because they are sure to suffer from same hurricanes during $T$ years. It is not same to the case that each turbine is located in a different wind farm with same surrounding. So Eq. (\ref{1q2}) can not take place of
 the state vector calculated by Eq. (\ref{Ef2}) completely.

\section{Concluding and remarks}
In this paper, the characters of hurricanes and wind turbines are discussed by similar methods as established by Rose {\it et. al} in \cite{Stephen2012}. In which the risk of wind turbine towers suffered from hurricanes is estimated by cumulative distribution function (CDF) of the number of  buckling turbine towers, and the hurricane risks in a wind farm are analyzed by the expected survived time (EST) of a single turbine. Monte Carlo simulations show that our results are accurate enough. The study in this paper is helpful to understand the effects of hurricanes on a wind farm, and may also be valuable to the design and maintenance of wind farms.


\newpage

\clearpage

\begin{table}
  \centering
  \begin{tabular}{l|l}
    Method&ESN\\\hline
    State Vector&5.8884\\
    Monte Carlo&5.8412\\
    Expected buckling probability&5.8885
  \end{tabular}
  \caption{The expected survive number (ESN) of 50 turbines totally in 20 years is calculated in three different ways. Turbines are actively yawing in Galveston County, TX. The parameter values used in these calculations are $\alpha=174$, $\beta=19.3$, $\lambda=0.19$, $\mu=78.7$, $\sigma=12.1$, $\xi=0.251$, see Eq. (\ref{GEV}), (\ref{loglogistic}) and (\ref{pb}).}\label{TestOfET}
\end{table}

\end{document}